\documentclass[runningheads]{llncs}

\usepackage{cite}
\usepackage{amsmath,amssymb,amsfonts}
\usepackage{algorithmic}
\usepackage{graphicx}
\usepackage{textcomp}
\usepackage[table,xcdraw]{xcolor}
\usepackage{enumitem}
\usepackage{multirow}
\usepackage{diagbox}
\usepackage{titlesec}
\usepackage{verbatim}
\usepackage{longtable} 
\def\BibTeX{{\rm B\kern-.05em{\sc i\kern-.025em b}\kern-.08em
    T\kern-.1667em\lower.7ex\hbox{E}\kern-.125emX}}

\titlespacing*{\section}{0pt}{12pt plus 4pt minus 4pt}{6pt plus 2pt minus 2pt}
\titlespacing*{\subsection}{0pt}{12pt plus 4pt minus 4pt}{6pt plus 2pt minus 2pt}

\begin{document}

% ----------------------------------------------------

\title{Trust Management in the Internet of Everything}%\\{\footnotesize Temporary title}


\author{Barbora Buhnova\inst{}}
\authorrunning{B. Buhnova}
\institute{Masaryk University, Faculty of Informatics \\ Brno, Czech Republic\\ 
\email{buhnova@mail.muni.cz}}

\maketitle

% ----------------------------------------------------

\begin{abstract}

Digitalization is leading us towards a future where people, processes, data and things are not only interacting with each other, but might start forming societies on their own. In these dynamic systems enhanced by artificial intelligence, trust management on the level of human-to-machine as well as machine-to-machine interaction becomes an essential ingredient in supervising safe and secure progress of our digitalized future.
This tutorial paper discusses the essential elements of trust management in complex digital ecosystems, guiding the reader through the definitions and core concepts of trust management. Furthermore, it explains how trust-building can be leveraged to support people in safe interaction with other (possibly autonomous) digital agents, as trust governance may allow the ecosystem to trigger an auto-immune response towards untrusted digital agents, protecting human safety.

% We will discuss the understanding of trust in traditional domains, such as psychology, sociology or economy, leading us to the emerging debates on how trust could be understood in the context of the Internet of Everything. We will go through the individual components of trust management, including trust assessment and its influencing factors (e.g., feeling vulnerable), trust propagation (e.g., via reputation), and trust updates (e.g., via dynamic certification schemes). We will give an overview of the current open discussions about the centralized/global vs decentralized/local trust management strategies and discuss their risks and related trust attacks (e.g, self-promotion or whitewashing).

% As the topic of trust governance in the context of dynamic autonomous ecosystems involving human-machine interplay is very recent, the interactive part of the tutorial will be organized as a working session around the most pressing research questions on the topic, possibly touching also the topics of ethical guarantees, privacy concerns and human-inherent subjectivity of trust.

\end{abstract}

% ----------------------------------------------------

\begin{keywords}
Trust Management, Internet of Everything, Autonomous Ecosystems, Software Architecture
\end{keywords}

% ----------------------------------------------------

\section{Introduction}
\label{sec:Introduction}

The Internet of Everything (IoE), which is constructed by networking people, processes, data and things (e.g., devices, appliances, vehicles)~\cite{farias2021internet}, is reshaping the vision of how humans and autonomous systems could engage in common collaborative goals. While this advanced digitalization opens up numerous opportunities for our sustainable future, it also imposes new safety concerns regarding the undesirable behavior of autonomous digital actors in these ecosystems (e.g., a driverless car intentionally causing dangerous road consequences).

% \paragraph{Trust as an essential ingredient}
% [REPHRASE - from abstract] In these highly unpredictable systems with a high degree of autonomy  enhanced by artificial intelligence, trust management on the level of human-to-machine as well as machine-to-machine interaction becomes an essential ingredient to supervise safe and secure progress of our digitalized future.

% -------------

\paragraph{Trustworthiness does not guarantee trust.}

Although approaches exist to ensure trustworthiness of the individual ecosystem components, via improving their security, reliability, availability, etc., trust is difficult to get addressed via such solutions due to the fact that trust is conceptually a belief about a system that is out of our control. Therefore, although the system might declare its trustworthiness, this does not give us any guarantee that it can be trusted. This is an effect of the fact that malicious agents can enter the ecosystem with the intention to disrupt the basic functionality of a network for malicious purposes while keeping their malicious intentions hidden behind announced collaborative goals~\cite{cioroaica2021goals, sagar2022understanding}. 

% -------------

\paragraph{Immune-response capabilities of trust-based ecosystems.}

One promising strategy in supporting the safe progress of human-machine partnership in the presence of malicious entities within IoE is based on involving the intelligence inside the individual ecosystem agents, which could be directed towards real-time detection of trust-breaking behaviour in other agents~\cite{Iqbal2022SMC} so that problematic agents are detected and isolated before they can engage in harmful behaviour. An effective trust-management system is, therefore, imperative for detecting and dealing with misbehaving agents before they jeopardize the ecosystem functionality~\cite{sagar2022understanding}.

% -------------

\paragraph{Scope of this paper.}

This paper serves as supporting material for a tutorial that gives an introduction to the main elements of trust management in complex digital ecosystems. We discuss the understanding of trust in traditional domains, such as psychology, sociology or economics, leading us to the emerging debates on how trust can be understood in the context of the Internet of Everything. We go through the individual components of trust management, including trust evaluation and its influencing factors (e.g., feeling vulnerable), trust propagation (e.g., via reputation), and trust updates (e.g., via dynamic certification schemes). We give an overview of the current open discussions about the centralized/global vs decentralized/local trust management strategies and discuss their risks and related trust attacks (e.g, self-promotion or whitewashing).

As the topic of trust governance in the context of dynamic autonomous ecosystems involving human-machine interplay is very recent, the paper also presents some insights from the interactive part of the tutorial, which was organized as a working session around the most pressing research questions on the topic, touching e.g. the topics of higher values in digital ecosystems (such as ethics or fairness), privacy concerns and human-inherent subjectivity of trust.

% ----------------------------------------------------

\section{Understanding Trust}
\label{sec:UnderstandingTrust}

While the research on trust in the context of Internet of Everything is quite recent, substantial body of knowledge exists in various research areas, namely discussing the definition of trust and approaches to building and managing trust. The main concepts relevant in the area of the Internet of Everything are outlined in this section.

% ------------------------

\begin{figure}[b!]
\centering
\includegraphics[width=\textwidth]{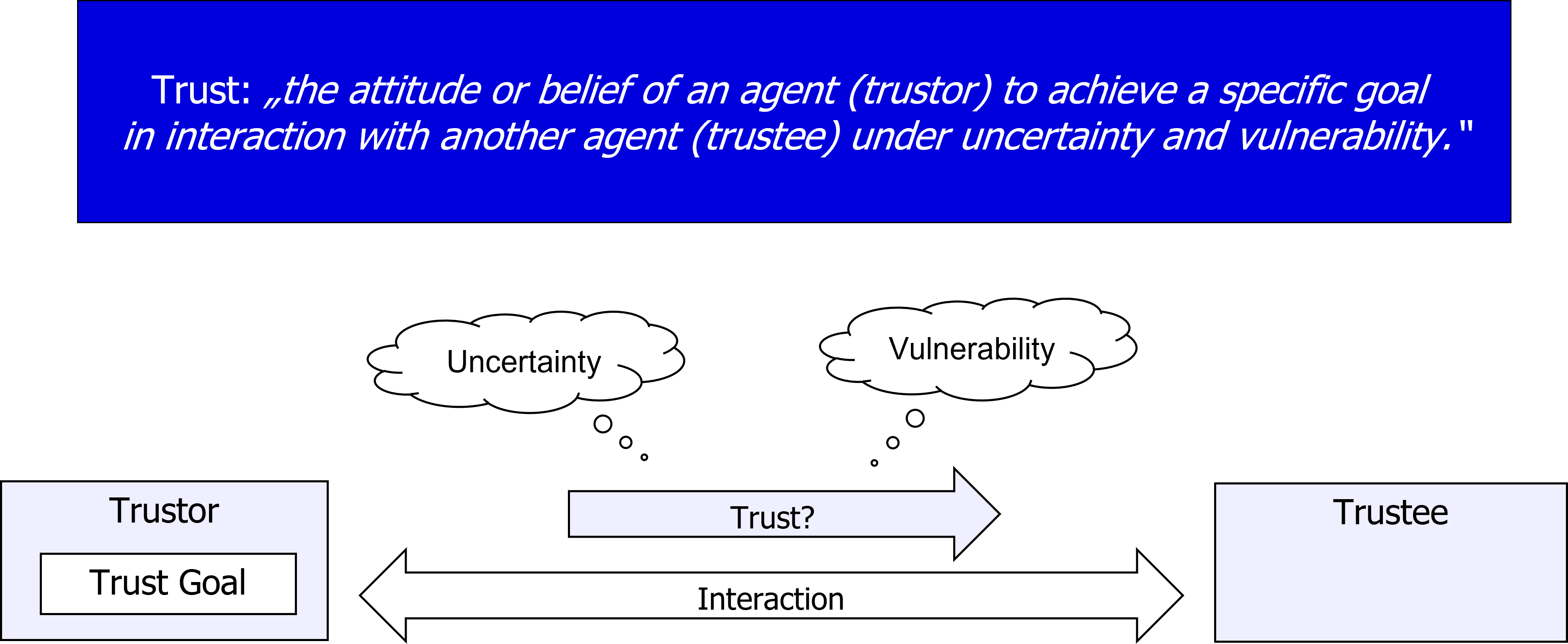}
\caption{Trust Definition for the Context of Internet of Everything (inspired from~\cite{lee2004trust}).}
\label{fig:TrustDefinition}
\end{figure}

\subsection{Definitions of Trust}

Trust is a complex phenomenon that has been studied in different contexts. This section outlines the definitions of trust in different domains.

\begin{itemize}

    \item \emph{Trust in Sociology:} Subjective probability that another party will perform an action that will not hurt my interest under uncertainty or ignorance~\cite{gambetta2000can}.\medskip
    
    \item \emph{Trust in Philosophy:} Risky action deriving from personal, moral relationships between two entities~\cite{lagerspetz1998trust}.\medskip
    
    \item \emph{Trust in Economics:} Expectation upon a risky action under uncertainty and ignorance based on the calculated incentives for the action~\cite{james2002trust}.\medskip
    
    \item \emph{Trust in Psychology:} Cognitive learning process obtained from social experiences based on the consequences of trusting behaviors~\cite{rotter1980interpersonal}.\medskip
    
    % \item \emph{Trust in Organizational Management:} Willingness to take risk and being vulnerable to the relationship based on ability, integrity, and benevolence~\cite{mayer1995integrative}.\medskip
    
    \item \emph{Trust in International Relations:} Belief that the other party is trustworthy with the willingness to reciprocate cooperation~\cite{kydd2007trust}.\medskip
    
    \item \emph{Trust in Automation:} Attitude or belief that an agent will help achieve another agent’s goal in a situation characterized by uncertainty and vulnerability~\cite{lee2004trust}.

    % Original: "Trust, a social psychological concept, seems particularly important for understanding human-automation partnerships. Trust can be defined as the attitude that an agent will help achieve an individual’s goals in a situation characterized by uncertainty and vulnerability~\cite{lee2004trust}."

    % My version I am using: "In the context of autonomous ecosystems, trust can be defined as the attitude or belief of an agent (trustor) to achieve a specific goal in interaction with another agent (trustee) under uncertainty and vulnerability~\cite{lee2004trust}."

\end{itemize}

In what follows, i.e. in IoE ecosystems, we adopt the definition inspired from the context of automation, extending the understanding of the agents to intelligent systems as well as humans and things, i.e. understanding trust as \emph{"the attitude or belief of an agent (trustor) to achieve a specific goal in interaction with another agent (trustee) under uncertainty and vulnerability"} (Figure~\ref{fig:TrustDefinition}).

% ------------------------

\subsection{Characteristics of Trust}

Trust in general is characterized by the following three characteristics~\cite{ghafari2020survey,sagar2022understanding}:

\begin{itemize}

    \item \emph{Subjective:} Trust is viewed using the centrality of an agent, wherein the trust is computed based on trustor’s observation (i.e., direct trust) as well as the opinion (i.e., feedback or indirect trust) of the other agents.\medskip
    
    \item \emph{Asymmetric:} Trust is an asymmetric property, i.e., if an agent A trusts another agent B, it does not guarantee that B also trusts A.\medskip
    
    \item \emph{Transitive:} System agent A is more likely to develop trust towards an agent B if A trust agent C that trusts agent B.
    
\end{itemize}

% ------------------------

\subsection{Scope of Trust Evaluation}

When evaluating trust, the scope of the considered context makes a difference. The scope that is typically considered is~\cite{ghafari2020survey,sagar2022understanding}:

\begin{itemize}

    \item \emph{Local:} It represents the trust based on an agent-agent relationship, wherein an agent evaluates the trustworthiness of another agent using local information such as its current observation and past experience.\medskip
    
    \item \emph{Global:} In comparison to the local trust, the global trust is considered as the reputation of an agent within the ecosystem, wherein the reputation each agent might be influenced by the local trust score of each of the other agents in the ecosystem.\medskip
    
    \item \emph{Context-specific:} Trust of an agent towards another agent varies with context. A trust relation between the agents is usually dynamic and depends on multiple factors such as temporal factors or location.

\end{itemize}

% ------------------------

\begin{figure}[b!]
\centering
\includegraphics[width=\textwidth]{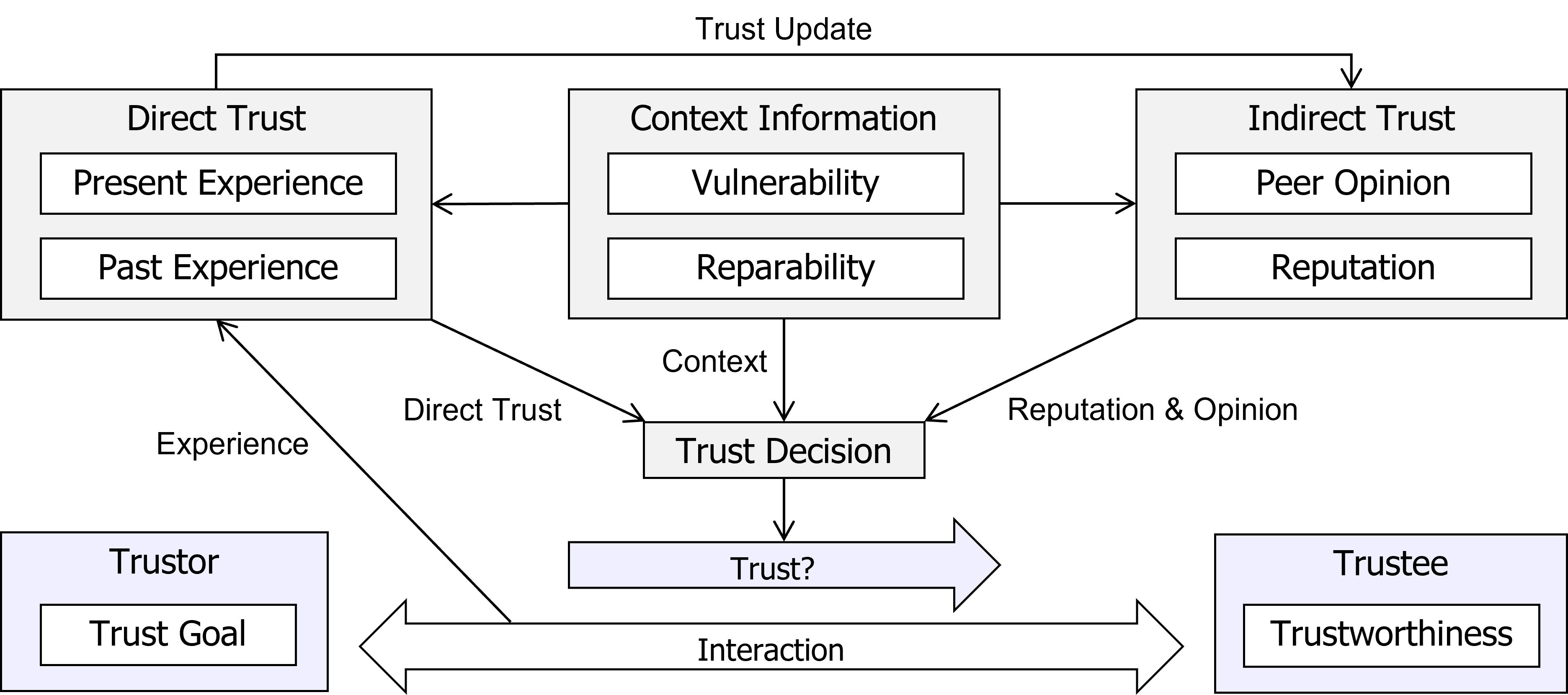}
\caption{Trust Management Components.}
\label{fig:TrustManagementComponents}
\end{figure}

\subsection{Trust Components}

Trust of one agent (trustor) in another (trustee) is being built via combining the mechanisms of direct trust (based on local and highly context-specific experience with the other agent) and indirect trust (possibly global reputation of the agent)~\cite{truong2017personal}. The two are later combined via the component of trust decision (see Figure~\ref{fig:TrustManagementComponents}).

\begin{itemize}

    \item \emph{Direct Trust} represents an individual judgment by the trustor from its direct interaction with the trustee. Specifically, it is being based on a combination of present and past experience (direct observation) of the trustor with the trustee. In this regard, mechanisms need to be in place to evaluate the experience during runtime interaction of the trustor with the trustee (possibly mimicking human cognitive processes) without exposing its vulnerabilities.\medskip
    
    \item \emph{Indirect Trust} is being inferred based on propagated opinions from different trust paths to judge the reputation of a trustee. The primary sources are thus the trusted peers of the trustor (their opinions and recommendation) and an authority overseeing the reputation of the trustee within the ecosystem. In this regard, mechanisms need to be in place to update the reputation and propagate it through the ecosystem.\medskip

    \item \emph{Context Information} influences the trust decision to reflect on the current situation of the trustor in which the trust decision is being made (e.g. how vulnerable it feels in interaction with the trustee, what risks it takes, whether the trustee will be held accountable in case of malicious behaviour, whether the possible harm is reversible or would be compensated, i.e. reparability is ensured). Besides, broader contextual information also influences the direct and indirect trust computation.

\end{itemize}

% ------------------------

\section{Trust Management Activities}
\label{sec:TrustManagementActivities}

The individual trust management activities are connected to the evaluation of the direct trust, management of the indirect trust, and the combination of the two in a trust decision~\cite{sagar2022understanding}.

% ------------------------

\subsection{Direct-Trust Evaluation}

The assessment of direct trust involves the computation of trust metrics characterizing the present/past experience together with aggregating the results into a trust score via trust formation.

% -------------

\paragraph{Trust Metrics} refer to the features that are chosen and combined in trust computation. These features can refer to:

\begin{itemize}

    \item \emph{QoS Metrics}, which represent the confidence that an agent is able to offer high quality of the delivered service, e.g. in terms of reliability, availability, security or accuracy~\cite{xiao2015guarantor}.\medskip
    
    \item \emph{Social Metrics}, which represent the social relationships among ecosystem agents, which can include integrity, benevolence, honesty, friendship, openness, altruism, or unselfishness~\cite{nitti2013trustworthiness,iqbal2023ENASE}.

\end{itemize}

% -------------

\paragraph{Trust Formation} forms the trust based on one of the following strategies~\cite{sagar2022understanding}:

\begin{itemize}

    \item \emph{Single aspect}, e.g., in terms of a positive or negative threshold for a specific trust metric.\medskip
    
    \item \emph{Multiple aspects}, i.e., more sophisticated trust model that includes various trust metrics~\cite{chen2015trust}.\medskip
    
    \item \emph{Aspect aggregation}, i.e. multiple aspects aggregated in a single trust score, for instance via a Weighted Sum, Belief Theory, Bayesian System, Fuzzy Logic, Regression Analysis, or Machine Learning~\cite{sagar2022understanding}.

\end{itemize}

% ------------------------

\subsection{Indirect-Trust Management}

The management of the indirect trust involves trust updates and trust propagation to ensure up-to-date opinion and reputation information that can feed trust decisions.

% -------------

\paragraph{Trust Update.}
At the end of a transaction or at any specified time interval, trust reputation of a trustee needs to be updated, which typically happens in one of the following ways:

\begin{itemize}

    \item \emph{Event-driven:} trust is updated after each transaction or once an event has occurred~\cite{chen2014trust}, which however increases the traffic overhead in the network.\medskip
    
    \item \emph{Time-driven:} trust is collected and updated periodically after a given interval of time~\cite{namal2015autonomic}.\medskip
    
    \item \emph{Hybrid:} trust is updated periodically and/or in case of an event (after an interaction)~\cite{xiao2015guarantor}.

\end{itemize}

% -------------

\paragraph{Trust Propagation.}
Trust propagation facilitates the understanding of how the trust propagates in the ecosystem and is generally categorized in~\cite{sagar2022understanding}:

\begin{itemize}

    \item \emph{Centralized schemes} rely on a centralized entity that is primarily responsible for (a) gathering trust-related information for the purpose of trust computation and (b) propagating it in the ecosystem~\cite{nitti2013trustworthiness}. Centralized controlled frameworks are vulnerable to a single point of failure, which can cause the entire ecosystem trust to collapse.\medskip
    
    \item \emph{Distributed schemes} rely on the individual agents being responsible for both trust computation and propagation within the ecosystem without any central authority~\cite{chen2014trust}. This scheme faces other inherent challenges, like honest trust computation, managing computational capabilities and unbiased trust propagation within the ecosystem.\medskip
    
    \item \emph{Hybrid schemes} are aimed at mitigating the challenges of the two earlier schemes, dividing the propagation into two categories, i.e., (1) locally distributed and globally centralized, and (2) locally centralized and globally distributed~\cite{nitti2013trustworthiness}.

\end{itemize}

% ------------------------

\subsection{Trust Decision} 

After computing the direct-trust score of a trustee and understanding its indirect-trust reputation (from the peers and/or central authority), the main purpose of a trust management system is to identify whether the trustee is to be considered trustworthy or untrustworthy by means of~\cite{sagar2022understanding}:

\begin{itemize}

    \item \emph{Threshold-based decision:} where the decision is taken on the basis of either a rank-based function or a (possibly dynamically changing) threshold value.\medskip
    
    \item \emph{Policy-based decision:} where more complex policies are used to identify and decide whether an agent is classified as trustworthy or not by using contextual information, e.g. in terms of location or temporal factors.

\end{itemize}

% ----------------------------------------------------

{\footnotesize
\begin{table}
\caption{Social Values in terms of Ethical Principles \cite{jobin2019global}}\label{tab:values}
\begin{tabular}{|p{.4cm}|p{2.5cm}|p{9cm}|}
\hline
1&Transparency &Transparency, explainability, explicability, understandability, interpretability, communication, disclosure, showing\\
\hline
2&Justice and fairness & Justice, fairness, consistency, inclusion, equality, equity, (non-)bias,
(non-)discrimination, diversity, plurality, accessibility, reversibility, remedy, redress, challenge, access and distribution\\
\hline
3&Non-maleficence & Non-maleficence, security, safety, harm, protection, precaution,
prevention, integrity (bodily or mental), non-subversion\\
\hline
4&Responsibility & Responsibility, accountability, liability, acting with integrity\\
\hline
5&Privacy &Privacy, personal or private information\\
\hline
6&Beneficence & Benefits, beneficence, well-being, peace, social good, common good\\
\hline
7&Freedom and autonomy&Freedom, autonomy, consent, choice, self-determination, liberty, empowerment\\
\hline
8&Trust & Trust\\
\hline
9&Sustainability &Sustainability, environment (nature), energy, resources (energy)\\
\hline
10&Dignity &Dignity\\
\hline
11&Solidarity &Solidarity, social security, cohesion\\
\hline
\end{tabular}
\end{table}
}

\section{Trust-Based Mechanisms for Ecosystem Wellbeing}
\label{sec:Wellbeing}

Once we have a way to assess the trustworthiness of the individual ecosystem members, we can employ the information in the actual support of the wellbeing of the ecosystem, e.g. in terms of its safety, fairness, solidarity, or other values (see Table~\ref{tab:values})~\cite{bangui2023ANTethical,jobin2019global}. Some examples of wellbeing-promoting mechanisms follow.

\paragraph{Incentives in Terms of Reward and Punishment.} One of the essential mechanisms for promoting ecosystem wellbeing is the ability of the ecosystem to incentivize behaviours that contribute to promoting the agreed values in the ecosystem. This is being done via the mechanisms of reward and punishment, which can be based e.g. on the actual trust score of the agent, or on the relative change (recent increase/decrease) of the trust score~\cite{bangui2023ANTindirecttrust, bangui2023SACdeeptrust}.

\paragraph{Evidence Collection for Justification and Reparation.} To ensure fairness in the ecosystem, mechanisms need to be in place that not only help us see there might be discrimination or unfairness happening in the system (e.g. newly joining agents having a hard time gaining sufficient trust needed to be allowed to participate fully in the ecosystem), but also help us justify decisions that might be opposed by certain ecosystem agents, or to correct trust misjudgment~\cite{bangui2023ANTindirecttrust}. Besides, evidence collection is a promising tool to detect trust attacks (discussed in Section~\ref{sec:TrustAttacks}) and hidden malicious intentions of ecosystem agents before they get fully revealed (discussed in Section~\ref{sec:Challenges}).

\paragraph{Safety Assurance in the Face of Untrusted Agents.} One of the essential ingredients of the \emph{immune-response capabilities of trust-based ecosystems} envisioned in Section~1 is the ability to isolate the misbehaving agents to protect the safety of the ecosystem. While this might be easy to do in the scenarios where trust navigates spreading of information or services (the trustor simply avoids using the knowledge and services by untrustworthy agents), it becomes challenging when trust shall navigate physical safety (e.g. to avoid collision with a malicious vehicle). In the latter case, an action by the authority overseeing agents' reputation might be needed, e.g., imposing restrictions of the agent function in effect of the reputation falling below certain threshold. In such cases, it is crucial to apply the mechanisms of adaptive function restriction to mitigate the risk of misjudging the trustworthiness of an agent~\cite{halasz2022fedcsis}. % Needless to say that there is also a risk of limited controllability of some of the agents, sometimes due to compatibility issues.

% ----------------------------------------------------

\begin{figure}[b!]
\centering
\includegraphics[width=0.8\textwidth]{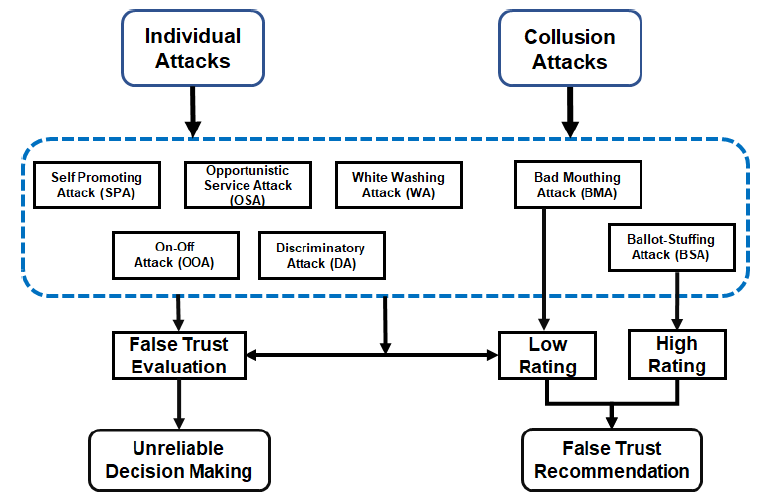}
\caption{Trust Attacks~\cite{sagar2022understanding}.}
\label{fig:TrustAttacks}
\end{figure}

\section{Trust Attacks}
\label{sec:TrustAttacks}

Given the fact that trust is largely influenced by the belief and perception of the individual ecosystem agents, there is a high risk of trust attacks, which can take different forms (see Figure~\ref{fig:TrustAttacks}), summarized below.

\paragraph{Individual Attacks:} refer to the attacks launched by an individual agent, which can take form of~\cite{sagar2022understanding,chahal2020trust}:

\begin{itemize}

    \item \emph{Self-Promoting Attacks:} In this type of attack, an agent promotes its significance by providing good recommendation for itself so as to be selected as a service provider, and then acts maliciously.\medskip
    
    \item \emph{Whitewashing Attacks:} In this attack, an agent exits and re-joins the ecosystem to recover its reputation and to wash-away its own bad reputation.\medskip
    
    \item \emph{Discriminatory Attacks:} In this type of attack, an agent explicitly attacks other agents that do not have common friends with it, i.e. it performs well for a particular service/agent and badly for some other services/agents.\medskip
    
    \item \emph{Opportunistic Service Attacks:} In this type of attack, an agent might offer a great service to improve its reputation when its reputation falls too low.\medskip
    
    \item \emph{On-Off Attacks:} In this type of attack, an agent provides good and bad services on and off (randomly) to avoid being labeled as a low-reputed agent.

\end{itemize}

% -------------

\paragraph{Collusion-based Attacks:} represent the attacks launched by a group of agents to either provide a high rating or low rating to a particular agent, such as~\cite{marche2020trust}:

\begin{itemize}

    \item \emph{Bad-Mouthing Attacks:} In this type of attack, a group of agents diminishes the reputation of a trustworthy agent within the ecosystem by providing bad recommendations about it.\medskip
    
    \item \emph{Ballot-Stuffing Attacks:} In this type of attack, a group of agents boosts the reputation of bad agents within the ecosystem by providing good recommendations for them.

\end{itemize}

% ----------------------------------------------------

\section{Scenarios}
\label{sec:Scenarios}

This section discusses examples of trust-management scenarios in the context of vehicular ecosystems, where trust management is crucial to avoid safety consequences of misbehaviour of individual smart agents.

% ------------------------

\begin{figure}[b!]
\centering
\includegraphics[width=0.7\textwidth]{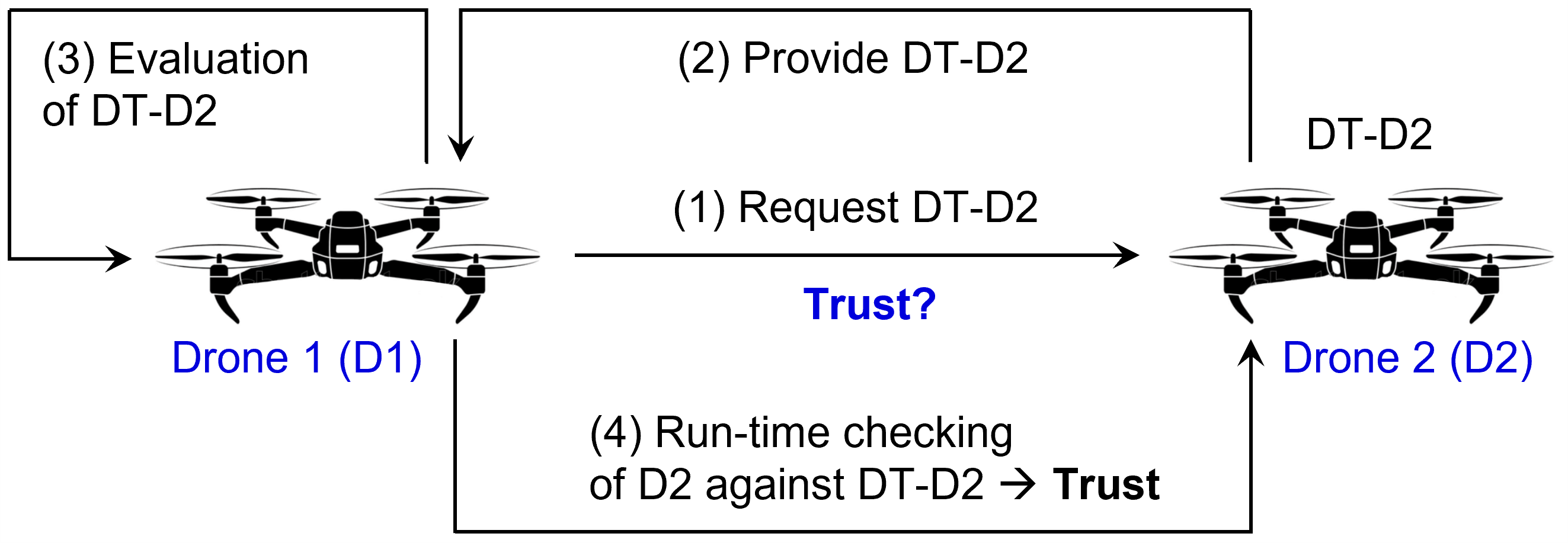}
\caption{Interaction Scenario~\cite{Iqbal2022SMC}.}
\label{fig:Drones}
\end{figure}

\subsection{Collision Avoidance with Misbehaving Vehicle}

Consider a situation of a vehicle (trustor) getting in the proximity of another vehicle (trustee), which could possibly be misbehaving (either due to a fault or being intentionally malicious) with an increased risk of causing a collision. In such a scenario, the possibility to collect present experience influencing the direct-trust evaluation is limited due to the risk of collision in case of very close interaction. A way to support the experience building in such a case, which we have explored in~\cite{Iqbal2022SMC}, is illustrated in Figure~\ref{fig:Drones} on an example of two drones. Consider Drone 1 (trustor) evaluating its direct trust in Drone 2 (trustee). The present-experience evaluation supporting the trust building in this scenario is formed in the following steps: First, Drone 1 asks Drone 2 to declare its intended behaviour in form of a Digital Twin (declared behaviour) that can be used by Drone 1 within the direct-trust evaluation process. In response, Drone 2 shares its Digital Twin (DT), which is checked by Drone 1 for its trustworthiness (absence of malicious actions) and then (if harmless), Drone 1 employs the DT in runtime compliance checking of the actual behaviour of Drone 2 against its Digital Twin (D2-DT). If at runtime Drone 2 deviates from its declared behaviour, Drone 1 can report Drone 2 to the ecosystem governance authority, which can internally decrease its reputation and start closer surveillance of Drone 2, possibly even employing some level of safeguarding of its operation (if the reputation falls too low or the associated threat gets too high).

% ------------------------

\subsection{Trust in Run-time Update Downloaded to a Vehicle}

The evolution in smart driving functions is pushing for frequent runtime updates being downloaded to the vehicles to improve their functionality, quality (e.g. vulnerability correction) or A/B test future software evolution. While numerous procedures are in place to ensure the trustworthiness of the software updates being downloaded to the vehicle, the risk is still there that these updates can contain vulnerabilities or hidden intended faults manifesting into malicious behavior~\cite{cioroaica2019not}. This can be caused even intentionally, with the help of an insider attacker engaged in the software update development. In this scenario, the vehicle shall have mechanisms in place to gradually build trust in the new software update before it allows it to control its critical driving functions, where both direct-trust evaluation and indirect-trust management can be integrated into the trust-building process. In~\cite{cioroaica2019not}, we have explored the potential of the Digital Twin of the update also in this scenario, to facilitate the experience gaining via predictive simulation of the update with the help of the Digital Twin.

% ------------------------

\subsection{Trust-based Vehicle Admission in a Platoon}

In the future automotive ecosystems, the formation of vehicle platoons is envisioned to support the traffic flow and reduced fuel consumption for more sustainable smart mobility. Consider a scenario where at the entry point of a highway, a vehicle can download a software solution that would assist it in platooning coordination~\cite{capilla2021autonomous}. When driving in a platoon, vehicles benefit from reduced fuel consumption due to reduced air friction, but the complex dynamics of collaborative and competitive forces could raise trust concerns about the safety of riding in such close proximity to other vehicles~\cite{cioroaica2021goals}. In this scenario, a vehicle in the platoon ecosystem can suddenly accelerate or decelerate and cause multiple car crashes within the platoon. Such a behavior can be caused by a fault or malicious intentions of the individual vehicle, or by a logic bomb in the downloaded software solution responsible for the platooning coordination, caused by a vulnerability inserted e.g. by a malicious insider working for the software provider. In this way, the solution can combine the two approaches outlined in the previous two scenarios.

% ------------------------

\subsection{Human-to-Vehicle Trust Building in Autonomous Driving}

The future automotive systems will be subject to a gradual process towards fully autonomous driving. During the process, human responsibilities will be gradually replaced by autonomous driving functions, at a speed governed by the level of trust of the humans in the autonomous driving capabilities of the vehicles. The acceptance of any new level of autonomy shall be built up gradually and governed by trust-management mechanisms, to ensure acceptance by both drivers and pedestrians. In~\cite{cioroaica2020building}, we envision the incremental development of autonomy on the road as a win-win situation for instrumenting the building of trust on both sides. In order to build a trusting relationship, the driver and the system shall in our view evolve together in a process that creates sustainable trust and supports the interchange between the human trustor with a smart software trustee.

% ----------------------------------------------------

\subsection{Information Exchange in Coordinated Ecosystem Moves}

An interesting scenario that came up in the tutorial discussion is linked to the high reliance on information exchange in coordinated ecosystem moves, e.g. when one vehicle is overtaking another one (whether driver-less or with a driver, possibly a cyclist) and needs to know its planned speed, when an emergency vehicle needs to be given priority when passing through a crossing, or when a human needs to be let to cross the street. Trust in this case needs to be built not only in the other agent in the ecosystem, but also in the actual information being exchanged, which could be influenced e.g. by a faulty sensor. In this regard, an interesting trust scenario is also between a vehicle and its own sensor, which might be faulty. And a great comment was given also about the individual pieces of information (e.g. that there is ice on the road in a particular location), which could act as the first-class IoE ecosystem entities with their own trust score governed by indirect trust management.

\section{Challenges of Trust Management in IoE}
\label{sec:Challenges}

Besides the trust attacks, which bring a considerable challenge in designing robust trust management systems, numerous other challenges of trust management in IoE have been discussed with the tutorial participants, and also outlined in literature~\cite{sagar2022understanding}. We summarize the key challenges below.

% -------------

\paragraph{Situational Scope of Trust.}

There is high context-dependence of trust, namely in terms of the location and time in their broader understanding~\cite{khan2020trust}. Besides, the peer network can change depending on the context of trust evaluation, similarly to the human context, where a person who can be trusted in the context of their own expertise, might be still untrustworthy in a field of knowledge where the person relies on the opinion of peers influenced by misinformation. This is also a reason why reputation shall be contextualized, as the reputation of an agent might be influenced by various dimensions of the context in which the trustworthiness is judged.

% -------------

\paragraph{Subjectivity of Trust.}

Trust is a highly subjective concept, being influenced not only by the context of the trustor, but also the individual trust goal of the trustor (the reason for which the trust is being established) and the factors inherent to the individual trustor (e.g. the willingness to take risk and trust despite the uncertainty, or willingness to rely on the peers recommendation). Furthermore, the peer network that influences the trust decision might itself be subjective or biased, which causes the propagation of subjectivity and bias in the ecosystem.

% -------------

\paragraph{Default Trust Score of New Agents.}

The current trust management solutions presume the initial trust score of a newly joined agent to be within the range $<0;0.5>$~\cite{sagar2022understanding}. Choosing the initial trust score (so-called cold start problem) is challenging on both sides of the interval. When the initial trust score is close to 0.5 (i.e. neutral, neither trustworthy nor untrustworthy)~\cite{chen2015trust}, this might lead to a malicious agent abusing its trust level to cause harm before it identifies as untrustworthy, or to white-wash its reputation by leaving and re-joining the ecosystem under new identity. On the other hand, when the initial trust score is close to 0, the new agent might never get the chance to engage with the rest of the ecosystem (i.e., not being trusted).

% -------------

\paragraph{Trust Erosion.}

Besides the fact that the trust score of any agent in the IoE ecosystem varies with time, its trust score (or reputation) is also subject to decay if there are no or too few interactions the agent is involved in~\cite{truong2017personal}. It is therefore important to consider the trust values lifespan where the trust score of inactive agents shall be subject to decay after a particular duration of time~\cite{sagar2022understanding}. 

% -------------

\paragraph{Detection of Hidden Malicious Intentions.}

As the overview of trust attacks above has shown, the malicious intentions of the IoE ecosystem agents can be very well hidden and hard to detect. Furthermore, when we employ very sensitive methods of detection, e.g. the discussed runtime compliance checking of the declared and actual trustee behaviour~\cite{Iqbal2022SMC}, we run into the risk of misjudging the trustworthiness of the agent, possibly reporting it as suspicious due to a slight deviation in the declared-actual behaviour although it might be well-behaving (with a deviation related to the accuracy of the declared behaviour). In that regard, it is crucial to have mechanisms in place for long-term evidence collection that can support richer reasoning on the behavioral patterns of the agents in the ecosystem, indicating e.g. the patterns resembling trust attacks or hiding the attempts of misbehaviour.

% -------------

\paragraph{Building Trust in the Trustworthy.}

Although much of the text has been centered around trust management directed towards detecting untrustworthy agents, it is equally important to explore methods supporting trust building in the trustworthy agents. Indeed, the fact that an agent is trustworthy does not guarantee it will be trusted by others, especially in face of trust attacks or biased peer opinion that might be directed against it. One of the promising techniques in promoting trust in this context is based on the ability of an agent to give evidence of its trustworthiness. In the context of AI-enhanced agents, for instance, this might be based on the explanation of agents' actions, which is shown to have positive influence on human participation and willingness to engage with the digital agents~\cite{garlan2020explanation}.

% -------------

\paragraph{High Degree of Dynamism and Uncertainty in IoE.}

As illustrated by different examples in this section, the dynamism and uncertainty inherent to the Internet of Everything is enormous. Agents can join and leave the ecosystem anytime, and they have very limited information about each other, facing high unpredictability of other agents and their environment. In such conditions, crucial information that is needed to make a trust decision might be missing, pushing trust models to include built-in mechanisms to deal with uncertainty and bias in the decisions.

% ----------------------------------------------------

\section{Conclusion}
\label{sec:Conclusion}

In the context of Internet of Everything (IoE), human and digital agents form coalitions for achieving collaborative or individual goals. In a constant risk of malicious attacks, ecosystem participants ask for strong guarantees of their collaborators’ trustworthiness, which can be achieved via the mechanisms of trust management. In this paper, we have presented the emerging research field of trust management in the context of IoE, discussing the definitions, concepts, examples of scenarios and research challenges relevant to the community of software architecture, which undergoes the transition from rather stable systems of systems towards dynamic autonomous ecosystems~\cite{capilla2021autonomous}.

% ----------------------------------------------------

\section*{Acknowledgement}

The work was supported by ERDF "CyberSecurity, CyberCrime and Critical Information Infrastructures Center of Excellence" (No. CZ.02.1.01/0.0/0.0/16\_019/ 0000822).

% \cite{cioroaica2019not, cioroaica2019towards, cioroaica2020building, cioroaica2021goals, cioroaica2022edi, cioroaica2022towards, capilla2021autonomous, Iqbal2022SMC, buhnova2022tutorial, bangui2023ANTethical, bangui2023ANTindirecttrust, bangui2023SACdeeptrust}

% ----------------------------------------------------

\begin{comment}
\section*{About the Speaker}

Barbora Buhnova is an Associate Professor and Vice-Dean at Masaryk University (MU), Faculty of Informatics (FI MU) in Brno. Following her research career in Germany and Australia, she now leads multiple research teams at Faculty of Informatics MU (software architecture) and Czech CyberCrime Centre of Excellence C4e (critical infrastructures), with special emphasis on trust management in autonomous ecosystems. She is the Steering Committee chair of the International Conference on Software Architecture (ICSA) and has been involved in organization of numerous leading conferences (e.g. in OC of ICSE, ESEC-FSE, ASE). She acts as a reviewer and (guest) editor in multiple journals (e.g. IEEE TSE, Springer EMSE, Elsevier SCP, Elsevier JSS, Springer SoSyM, Wiley SME), and is member of the IEEE TSE Review Board. Next to her academic activities, she is the chair of the Association of Industrial Partners at FI MU (with 30+ companies), and is a Co-Founding and Governing Board member of Czechitas, a non-profit organisation aiming at making IT skills more accessible to youth and women (with 30,000+ graduates).

\end{comment}

% ----------------------------------------------------

\bibliographystyle{splncs04}
\bibliography{trust.bib}

\end{document}